\documentclass{elsart}
 \usepackage{epsfig}
 \usepackage{amssymb}
 \usepackage{amsmath}

\begin{document}
\begin{frontmatter}

\begin{flushright}
{INFNFE-01-01}
\end{flushright}

\title
{Atmospheric neutrino flux supported by recent muon experiments}

\author[1]{Giovanni Fiorentini\thanksref{EmGF}},
\thanks[EmGF]{E-mail: fiorentini@fe.infn.it}
\author[1,2]{Vadim A. Naumov\corauthref{cor}\thanksref{Supp}},
\corauth[cor]{Corresponding author.
              E-mail: naumov@fe.infn.it, naumov@api.isu.runnet.ru}
\thanks[Supp]{Partially supported by grant from Ministry of Education
              of Russian Federation, within the framework of the
              program ``Universities of Russia -- Basic Researches'',
              grant No.015.02.01.004.}
\author[1]{Francesco L. Villante\thanksref{EmFV}}
\thanks[EmFV]{E-mail: villante@fe.infn.it}
 \address[1]{Dipartimento di Fisica and Sezione INFN di Ferrara,
             Via del Paradiso 12, I-44100 Ferrara, Italy}
%\address[1]{Dipartimento di Fisica, Universit\`{a} di Ferrara
%            and INFN, Sezione di Ferrara Via del Paradiso 12,
%            I-44100 Ferrara, Italy}
 \address[2]{Laboratory for Theoretical Physics, Irkutsk State
             University, Gagarin boulevard 20, RU-664003 Irkutsk,
             Russia}

\begin{abstract}
       We present a new one-dimensional calculation of low and
       intermediate energy atmospheric muon and neutrino fluxes,
       using up-to-date data on primary cosmic rays and hadronic
       interactions. We study several sources of uncertainties
       relevant to our calculations.  A comparison with the muon
       fluxes and charge ratios measured in several modern
       balloon-borne experiments suggests that the atmospheric
       neutrino flux is essentially lower than one used for the
       standard analyses of the sub-GeV and multi-GeV neutrino
       induced events in underground detectors.
\end{abstract}
\end{frontmatter}

\newpage

%%%%%%%%%%%%%%%%%%%%%%%%%%%%%%%%%%%%%%%%%%%%%%%%%%%%%%%%%%%%%%%%%%%%%
\protect\section{Introduction}
\label{sec:Introduction}
%%%%%%%%%%%%%%%%%%%%%%%%%%%%%%%%%%%%%%%%%%%%%%%%%%%%%%%%%%%%%%%%%%%%%

The atmospheric neutrino (AN) flux is one of the main tools for
studying neutrino oscillations, neutrino decay and nonstandard
neutrino-matter interactions. Furthermore it represents an
unavoidable background for key experiments in astroparticle physics,
as those searching for proton decay, $n-\overline{n}$ transitions in
nuclei, etc. Therefore, the accuracy of theoretical predictions of
the AN flux ultimately affects the interpretation of many
experimental data.
%% and consistency of conclusions concerning some ``new physics''.

After several studies performed in the frameworks of
one-dimensional (1D)~%
\cite{BN1,Bartol,Honda} and three-dimensional (3D)~%
\cite{Lee90,FLUKA,Tserkovnyak99} cascade models, the physics of
neutrino production in the sub- and multi-GeV energy ranges is
fairly well understood.
%%%%%%%%%%%%%%%%%%%%%%%%%%%%%%%%%%%%%%%%%%%%%%%%%%%%%%%%%%%%%%%%%%%%%
%%% (see refs. \cite{Gaisser96,Lipari00,Battistoni01} for reviews
%%% and further references, and also the up-to-the-minute paper
%%% \cite{Plyaskin01} with rather unusual results).
%%%%%%%%%%%%%%%%%%%%%%%%%%%%%%%%%%%%%%%%%%%%%%%%%%%%%%%%%%%%%%%%%%%%%
However, the validity of the AN flux calculations is still
controversial. This is mainly related to the uncertainties in the
required input data (inclusive cross sections for particle
production, primary cosmic ray spectrum and composition, etc.).
Moreover, the problem is too complex to be solved without
many simplifications (explicit or implicit) and these also are
essential sources of uncertainties in the predicted AN fluxes
\cite{Gaisser96,Lipari00,Battistoni01,Plyaskin01}.
A popular opinion is that the accuracy of the present-day
calculations of the low-energy AN flux is about 20--30\%.
Sometimes this belief of obscure origin is treated as the uncertainty
in the {\em normalization factor}. But, as a matter of fact, the
uncertainties in the input data are so large that the actual accuracy
of the AN flux turns out to be much worse. It is also essential that
the uncertainties are very energy dependent and therefore they cannot
be reduced to a simple renormalization.

%%%%%%%%%%%%%%%%%%%%%%%%%%%%%%%%%%%%%%%%%%%%%%%%%%%%%%%%%%%%%%%%%%%%%
%%% It is also essential that the uncertainty grows with decreasing
%%% neutrino energy and therefore cannot be reduced to a simple
%%% renormalization.
%%%%%%%%%%%%%%%%%%%%%%%%%%%%%%%%%%%%%%%%%%%%%%%%%%%%%%%%%%%%%%%%%%%%%

In this work we present a new 1D calculation of low and intermediate
energy muon and neutrino fluxes, based on up-to-date data on primary
cosmic-ray flux and hadronic interactions. Our calculation is based
on a kinetic approach. In order to check the validity of our
description of hadronic interactions and shower development, we
perform a comparison of the predicted atmospheric muon fluxes with
the data from several recent balloon-borne experiments.

We believe that an accurate 1D AN flux calculation is still helpful,
despite the results of 3D calculations are now available.
Indeed, the main lessons we got from the recent 3D analyses~%
\cite{FLUKA,Tserkovnyak99,Lipari00,Battistoni01} may be
summarized as follows.
\begin{itemize}
\item 3D effects lead to a strong modification of zenith and
      azimuth angle distributions of neutrinos at energies
      $E_\nu\lesssim0.5$~GeV. On the other hand, the corresponding
      effect for the neutrino-induced events in underground detectors
      is expected to be modest and scarcely measurable, because their
      distributions are close to isotropic out of a very big mean
      scattering angle in quasielastic neutrino-nucleus collisions at
      low energies.
%%%%%%%%%%%%%%%%%%%%%%%%%%%%%%%%%%%%%%%%%%%%%%%%%%%%%%%%%%%%%%%%%%%%%
%%%   Besides, the correct interpretation of the measured angular
%%%   distributions of the low-energy $\nu$ induced particles is
%%%   hampered by essential experimental and theoretical
%%%   uncertainties in the relevant double-differential cross
%%%   sections.
%%%%%%%%%%%%%%%%%%%%%%%%%%%%%%%%%%%%%%%%%%%%%%%%%%%%%%%%%%%%%%%%%%%%%
\item Even at very low energies the 3D corrections to the $4\pi$ and
      even $2\pi$ averaged AN fluxes are relatively small. At least,
      they are comparable with the uncertainties in the input
      data, including the uncertainties in the cross sections for
      exclusive neutrino-nucleus interactions needed for simulating
      the neutrino-induced events in the underground detectors.
\item Above $1-2$~GeV the 3D effects become negligible for the
      neutrino fluxes averaged over the azimuth angles at any
      zenith angle.
\end{itemize}

On that ground, we expect the calculation offered for consideration
to be valid without restrictions at $E_\nu\gtrsim1$~GeV and we
consider it as a basis for future 3D generalization of our approach
to the AN problem.

Furthermore, our 1D code is useful for testing 1D and 3D codes with
a method {\em completely different from Monte Carlo}. It can be used
to perform a systematic analysis of uncertainties in the AN flux
calculations
%%%%%%%%%%%%%%%%%%%%%%%%%%%%%%%%%%%%%%%%%%%%%%%%%%%%%%%%%%%%%%%%%%%%%
%%% (including a verification of particle interaction models and
%%% models for primary cosmic-ray spectrum and composition)
%%%%%%%%%%%%%%%%%%%%%%%%%%%%%%%%%%%%%%%%%%%%%%%%%%%%%%%%%%%%%%%%%%%%%
and to adjust some poorly known input parameters, using the data
on cosmic-ray secondaries. In this work, in order to evaluate
quantitatively the uncertainties related to the particle
interactions,
%%%%%%%%%%%%%%%%%%%%%%%%%%%%%%%%%%%%%%%%%%%%%%%%%%%%%%%%%%%%%%%%%%%%%
%%% we consider different models for meson production in
%%% nucleon-nucleus and nucleus-nucleus collisions.
%%%%%%%%%%%%%%%%%%%%%%%%%%%%%%%%%%%%%%%%%%%%%%%%%%%%%%%%%%%%%%%%%%%%%
we perform a comparison of the results obtained with different
models for meson production in nucleon-nucleus and nucleus-nucleus
collisions.

%%%%%%%%%%%%%%%%%%%%%%%%%%%%%%%%%%%%%%%%%%%%%%%%%%%%%%%%%%%%%%%%%%%%%
\protect\section{The CORT code}
\label{sec:CORT}
%%%%%%%%%%%%%%%%%%%%%%%%%%%%%%%%%%%%%%%%%%%%%%%%%%%%%%%%%%%%%%%%%%%%%

The present work is based on an updated code {CORT}
({\sf C}osmic-{\sf O}rigin {\sf R}adiation {\sf T}ransport),
previously developed in \cite{BN2} and used in \cite{BN1} (see
also \cite{BN3,Bugaev98}). Like the earlier version, the current
Fortran 90 code implements a numerical solution of a system of 1D
kinetic equations describing the propagation of nuclei, nucleons,
light mesons, muons and (anti)neutrinos of low and intermediate
energies through a spherical, nonisothermal atmosphere. It takes into
account solar modulation and geomagnetic effects, energy loss of
charged particles, muon polarization and depolarization effect. The
exact kinematics is used in description of particle interactions and
decays. Let us sketch some most distinctive aspects.

In order to evaluate geomagnetic effects and to take into
account the anisotropy of the primary cosmic-ray flux in the
vicinity of the Earth, we use the method of ref. \cite{BN2}
and detailed maps of the effective vertical cutoff rigidities
from ref. \cite{Dorman71}.%
\footnote{The penumbral effects are included in the definition
          of the effective cutoff.}
The maps are corrected for the geomagnetic pole drift and compared
with the later results reviewed by Smart and Shea \cite{Smart94} and
with the recent data on the proton flux in near earth orbit obtained
with the Alpha Magnetic Spectrometer (AMS) \cite{AMS}.
The interpolation between the reference points of the maps is
performed by means of two-dimensional local B-spline.
The Quenby-Wenk relation (see e.g. \cite{Dorman71}),
re-normalized to the vertical cutoffs, is applied for evaluating
the effective cutoffs for oblique directions.
More sophisticated effects, like the short-period variations of the
geomagnetic field, Forbush decrease, re-entrant cosmic-ray albedo
contribution, etc., are neglected. We also neglect the geomagnetic
bending of the trajectories of charged secondaries and multiple
scattering effects.
Validity of our treatment of secondary nucleons and nuclei was
confirmed using all available data on proton spectra in the
atmosphere (see \cite{BN2}).

The meteorological effects are included using the Dorman model of the
atmosphere \cite{Dorman72} which assumes an isothermal stratosphere
and constant gradient of temperature (as a function of depths) below
the tropopause.  Ionization, radiative and photonuclear muon energy
losses are treated as continuous processes.
This approximation is quite
tolerable for atmospheric depths $h\lesssim2\times10^3$~g/cm${}^2$ at
all energies of interest \cite{Naumov94}.  Propagation of $\mu^+$ and
$\mu^-$ originating from every source (pion or kaon decay) is
described by separate kinetic equations for muons with definite
polarization at production. These equations automatically
account for muon depolarization through the energy loss (but not
through the Coulomb scattering).

In the new version, essentially all approximate formulas used
in \cite{BN1,BN2,BN3,Bugaev98} are replaced (or duplicated for a
control) with the more accurate numerical equivalents.
%%%%%%%%%%%%%%%%%%%%%%%%%%%%%%%%%%%%%%%%%%%%%%%%%%%%%%%%%%%%%%%%%%%%%
%%% The code has a number of options and run modes; it is rather
%%% flexible and permits modification of many input data with
%%% intrinsic switch keys.
%%%%%%%%%%%%%%%%%%%%%%%%%%%%%%%%%%%%%%%%%%%%%%%%%%%%%%%%%%%%%%%%%%%%%

%%%%%%%%%%%%%%%%%%%%%%%%%%%%%%%%%%%%%%%%%%%%%%%%%%%%%%%%%%%%%%%%%%%%%
\protect\section{Primary cosmic ray spectrum and composition}
\label{sec:Primaries}
%%%%%%%%%%%%%%%%%%%%%%%%%%%%%%%%%%%%%%%%%%%%%%%%%%%%%%%%%%%%%%%%%%%%%

In the present calculations, the nuclear component of primary cosmic
rays is broken up into 5 principal groups: H, He, CNO, Ne-S and Fe
with average atomic masses $A$ of 1, 4, 15, 27 and 56, respectively.
We do not take into account the isotopic composition of the primary
nuclei and assume $Z=A/2$ for $A>1$, since the expected effect on the
secondary lepton fluxes is estimated to be small with respect to
present-day experimental uncertainties in the absolute cosmic-ray
flux and chemical composition.%
\footnote{However, a more rigorous analysis must allow at least for
          small fractions of primary ${}^2$H and ${}^3$He affecting
          the $\mu^+/\mu^-$ and $\nu/\overline{\nu}$ ratios.}

We use the following 9-parameter model for the differential energy
spectra of every nuclear group at the top of the atmosphere
\begin{equation}\label{PCRS}
\frac{\d F_{\mathrm{A}}(E)}{\d E}=\left\{
\begin{matrix}
F_{\mathrm{A}}^1\epsilon_{\mathrm{A}}^%
{-\Gamma_{\mathrm{A}}(\epsilon_{\mathrm{A}})}
& \mbox{at} & E<W_{\mathrm{A}},      \\
F_{\mathrm{A}}^2\epsilon_{\mathrm{A}}^{-\gamma_{\mathrm{A}}}
& \mbox{at} & E>W_{\mathrm{A}},
\end{matrix}
\right.
\end{equation}
with
\begin{equation}\label{Gamma}
\Gamma_{\mathrm{A}}(\epsilon_{\mathrm{A}})=
\sum_{k=0}^3\gamma_{\mathrm{A}}^{(k)}\ln^k\epsilon_{\mathrm{A}},
\quad
\epsilon_{\mathrm{A}}=
{r_{\mathrm{A}}E}/{W_{\mathrm{A}}}.
%%%%%%%%%%%%%%%%%%%%%%%%%%%%%%%%%%%%%%%%%%%%%%%%%%%%%%%%%%%%%%%%%%%%%
%%% \quad \mathrm{A} = \mathrm{H},\mathrm{He},\ldots.
%%%%%%%%%%%%%%%%%%%%%%%%%%%%%%%%%%%%%%%%%%%%%%%%%%%%%%%%%%%%%%%%%%%%%
\end{equation}
Here $E$ is the total energy in GeV/nucleon.
The condition for continuity of the spectrum and its first
derivative in the point $E=W_{\mathrm{A}}$ yields
\[
\ln\left(F_{\mathrm{A}}^2/F_{\mathrm{A}}^1\right)=
\sum_{k=1}^3k\gamma_{\mathrm{A}}^{(k)}\ln^{k+1}r_{\mathrm{A}},
\quad
\gamma_{\mathrm{A}}=\sum_{k=0}^3(k+1)
\gamma_{\mathrm{A}}^{(k)}\ln^kr_{\mathrm{A}}.
\]
Thus, we remain with 7 independent parameters for each group which we
determine by fitting the experimental data.

For the hydrogen and helium groups at $E<120$~GeV/nucleon we chose
the data of the balloon-borne experiment BESS obtained by a flight in
1998 \cite{Sanuki00} (a period close to a minimum of solar activity).
For higher energies (but below the knee) we use data by a series
of twelve balloon flights of JACEE \cite{Asakimori98} and the result
of an analysis by Wiebel-Sooth et al. \cite{Wiebel-Sooth98} based
upon a very representative compilation of world data on primaries and
solid theoretical considerations. Since the fits obtained in
\cite{Wiebel-Sooth98} turned out to be very close to the combined
result
of the JACEE experiments, we call our model ``BESS+JACEE'' fit. A
maximum likelihood analysis with $W_1=W_4=2$~TeV/nucleon gives the
following values for the remaining parameters in eqs.~(\ref{PCRS})
and (\ref{Gamma}):
\[
\begin{aligned}
F_1^1          &= 5.7216\!\times\!10^{-5}, &\quad
F_4^1          &= 1.6500\!\times\!10^{-7},  \\
F_1^2          &= 5.8511\!\times\!10^{-5}, &\quad
F_4^2          &= 1.7239\!\times\!10^{-7},  \\
r_1            &= 1.5952,                  &\quad
r_4            &= 0.4783,                   \\
\gamma_1       &= 2.800,                   &\quad
\gamma_4       &= 2.655,                    \\
\gamma_1^{(0)} &= 2.7119,                  &\quad
\gamma_4^{(0)} &= 2.7881,                   \\
\gamma_1^{(1)} &= 6.9771\!\times\!10^{-2}, &\quad
\gamma_4^{(1)} &= 1.1953\!\times\!10^{-1},  \\
\gamma_1^{(2)} &= 3.2396\!\times\!10^{-2}, &\quad
\gamma_4^{(2)} &= 2.8694\!\times\!10^{-2},  \\
\gamma_1^{(3)} &= 3.8561\!\times\!10^{-3}, &\quad
\gamma_4^{(3)} &= 2.0829\!\times\!10^{-3},
\end{aligned}
\]
where $F_{\mathrm{A}}^i $ are in
m${}^{-2}$s${}^{-1}$sr${}^{-1}$(GeV/nucleon)${}^{-1}$. Clearly
the many digits in the above parameters are by no means indicating
the accuracy of the data, but they are necessary to avoid loss of
accuracy in numerical calculations.

\begin{figure}[htb]
\center\mbox{\epsfig{file=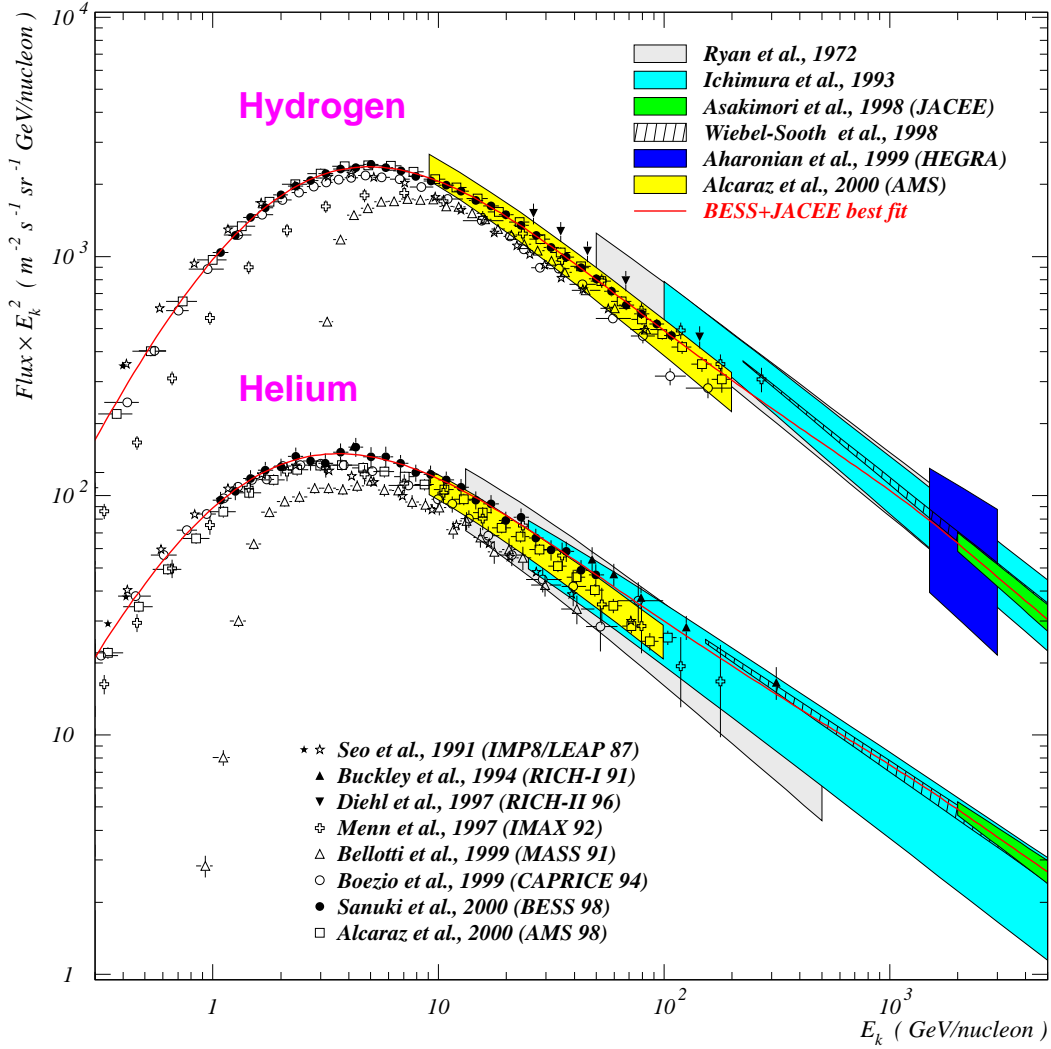,width=13.92cm}}
\protect\caption{Differential proton and helium spectra as a function
                 of kinetic energy per nucleon.
                 The data points and filled/shaded areas are from~%
                 \cite{Sanuki00,Asakimori98,Wiebel-Sooth98,%
                       Ryan72,Seo91,Ichimura93,Buckley94,Diehl97,%
                       Menn97,Aharonian99,Bellotti99,Boezio99a,%
                       Alcaraz00b,Alcaraz00c}.
                 Solid curves represent the BESS+JACEE best fit
                 (see text).
\label{f:PS}}
\end{figure}

Figure~\ref{f:PS} shows a comparison between the BESS+JACEE fit,
the data from \cite{Sanuki00,Asakimori98}, the fit
from \cite{Wiebel-Sooth98} (shaded areas) and several other
experiments~%
\cite{Ryan72,Seo91,Ichimura93,Buckley94,Diehl97,Menn97,%
      Aharonian99,Bellotti99,Boezio99a,Alcaraz00b,Alcaraz00c},
performed in different periods of solar activity. The filled/shaded
areas in fig.~\ref{f:PS} represent the power-type parametrizations
derived in%
\cite{Asakimori98,Ryan72,Ichimura93,Aharonian99,Alcaraz00b,%
      Alcaraz00c}
from the original data. The legend indicates the publication dates and
(when relevant) the years of measurements.
%--------------------------------------------------------------------
%--------------------------------------------------------------------

One sees that the BESS+JACEE fit is in excellent agreement with the
recent AMS data on the proton flux \cite{Alcaraz00b}. On the other
hand it slightly exceeds the AMS helium flux \cite{Alcaraz00c}
(with average discrepancy of about 15\%).
%% above 300~MeV/nucleon the average discrepancy is about 15\% and
%% increases with energy at $E\gtrsim30$~GeV/nucleon.
Our choice of the BESS\,98 data was in particular conditioned by the
fact that the power-law extrapolation of the high-energy tail
of the AMS helium rigidity spectrum ($\propto R^{-(2.74\pm0.02)}$)
\cite{Alcaraz00c} distinctly worse joins the world-averaged fit
($\propto E^{-(2.64\pm0.02)}$) \cite{Wiebel-Sooth98}
and even the JACEE\,1-12 fit
($\propto E^{-\left(2.68^{+0.04}_{-0.06}\right)}$)
\cite{Asakimori98}.%
\footnote{This is all the more true for the remaining data presented
          in fig.~\ref{f:PS}.}

We assume that the spectra of the remaining three nuclear groups are
similar to the helium spectrum,
\[
\frac{\d F_{\mathrm{A}}(E)}{\d E}=
c_A\frac{\d F_{\mathrm{4}}(E)}{\d E}
\quad (\mathrm{A}=15,27,56).
\]
This assumption does not contradict the world data for the CNO and
Ne-S nuclear groups but works a bit worse for the iron group.
Nevertheless, a more sophisticated model would be unpractical since
the corresponding correction would affect the secondary lepton fluxes
by a negligible margin. A normalization using the relevant data
yields $c_{15}=0.068$, $c_{27}=0.026$, $c_{56}=0.0054$.

In this paper we do not consider the effects of solar modulation and
use the BESS+JACEE spectrum in all calculations. Therefore the
predicted muon and neutrino fluxes are to some extent the maximum
ones possible within our 1D cascade model.

%%%%%%%%%%%%%%%%%%%%%%%%%%%%%%%%%%%%%%%%%%%%%%%%%%%%%%%%%%%%%%%%%%%%%
\protect\section{Nucleon-nucleus and nucleus-nucleus interactions}
\label{sec:interactions}
%%%%%%%%%%%%%%%%%%%%%%%%%%%%%%%%%%%%%%%%%%%%%%%%%%%%%%%%%%%%%%%%%%%%%

Calculations with the earlier version of
CORT \cite{BN1,BN2,BN3,Bugaev98} were based on semiempirical models
for inclusive nucleon and light meson production in collisions of
nucleons with nuclei, proposed by Kimel' and Mokhov (KM)
\cite{Kimel'74} and by Serov and Sychev (SS) \cite{Serov73} (see also
\cite{Kalinovsky85,Sychev99} for the most recent versions). The KM
model is valid for projectile nucleon momenta above $\sim4$~GeV/c and
for the secondary nucleon, pion and kaon momenta above 450, 150 and
300~MeV/c, respectively. Outside these ranges (that is mainly within
the region of resonance production of pions) the SS model was used.
Comparison of the KM model with the data and with some other models
is discussed in \cite{Gaisser96,Naumov93-00}.

Both models are in essence comprehensive parametrizations of the
relevant accelerator data. In our opinion, the combined ``KM+SS''
model provides a rather safe and model-independent basis to the
low-energy atmospheric muon and neutrino calculations. However it is
not free of uncertainties. For the present calculation, the fitting
parameters of the KM model for meson and nucleon production off
different nuclear targets were updated using accelerator
data not available for the original analysis
\cite{Kimel'74,Kalinovsky85}. The values of the parameters were
extrapolated to the air nuclei (N, O, Ar, C).%
\footnote{The old version of CORT used the renormalized $N$Be
          inclusive cross sections.}
The overall correction is less than 10-15\% within the kinematic
regions significant to atmospheric cascade calculations.  Besides,
energy-dependent correction factors were introduced into the model to
tune up the output $\pi^+/\pi^-$ ratio taking into account the
relevant new data. This correction only slightly affects the total
meson multiplicities and modifies the pion charge ratio within 5-9\%.

The processes of meson regeneration and charge exchange
($\pi^\pm+\mathrm{Air}\to\pi^{\pm(\mp)}+X$ etc.) are not of critical
importance for production of leptons with energies of our interest
and can be considered in a simplified way. Here we use a proper
renormalization of the meson interaction lengths, which was deduced
from the results of ref. \cite{Vall86} obtained for high-energy
cascades.

The next important ingredient of any cascade calculations is a model
for nucleus-nucleus collisions. The overwhelming majority of muon and
neutrino flux predictions is based on the simplest superposition
model (SM) in which the collision of a nucleus A, with total energy
$E_{\mathrm{A}}$ and atomic weight $A$, against a target nucleus B,
is treated as the superposition of $A$ independent collisions of
nucleons with the target nucleus, each nucleon having an energy
$E_0=E_{\mathrm{A}}/A$. This approximation is based on the hypothesis
that, when the energy per nucleon of the projectile is much larger
than the single nucleon binding energy, the $A$ nucleons interact
incoherently.  Generally speaking, this approach is rather far from
reality and its applicability to the lepton flux calculations is not
evident. A more accurate treatment of nucleus-nucleus collisions
becomes especially important at low energies and for the regions or
directions with high geomagnetic cutoffs. Indeed, the magnetic
rigidity of a proton bounded in a nucleus is a factor $A/Z$ larger
than that of a free proton of the same energy and therefore
(considering the steepness of the primary spectrum) mainly nuclei are
responsible for production of the low-energy secondaries.

Here we consider a modest generalization of a simple
``Glauber-like'' model used in \cite{BN1,BN2,BN3}.
Namely, we write the inclusive spectrum of secondary particles
$c$ ($c=p$, $n$, $\pi^\pm$, $K^\pm$, $K^0,\ldots$) produced in AB
collisions as
\begin{align}\label{AB->cX}
\frac{\d N_{\mathrm{AB} \to cX}}{\d x} = & \xi_{\mathrm{AB}}^c
\left[Z \frac{\d N_{p\mathrm{B} \to cX}}{\d x}+
   (A-Z)\frac{\d N_{n\mathrm{B} \to cX}}{\d x}\right]\nonumber\\
&+\left(1-\xi_{\mathrm{AB}}^c\right)
\left[Z\delta_{cp}+(A-Z)\delta_{cn}\right]\delta(1-x).
\end{align}
Here $\d N_{N\mathrm{B} \to cX}/\d x$ is the spectrum of particles
$c$ produced in $N\mathrm{B}$ collisions ($N=p,n$) and
$\xi_{\mathrm{AB}}^c$ is the average fraction of inelastically
interacting nucleons of the projectile nucleus A. The term
proportional to delta function describes the contribution of
``spectator'' nucleons from the projectile nucleus.

In the standard Glauber-Gribov multiple scattering theory the
quantity $\xi_{\mathrm{AB}}^c$ is certainly independent of the type
of inclusive particle $c$. On the other hand, it depends of the
type of nucleus collision. Indeed, essentially all nucleons
participate in the central AB collisions
($\xi_{\mathrm{AB}}^c\simeq1$)%
\footnote{Here we suppose for simplicity that the atomic weight of
          the projectile nucleus is not much larger than that of
          the target nucleus.}
while, according to the well-known Bialas--Bleszy\'nski--Czy\.z
(BBC) relation \cite{Bialas76},
\begin{equation}\label{BBC}
\xi_{\mathrm{AB}}^c=\sigma_{N\mathrm{B}}^{\mathrm{inel}}/
                    \sigma_{\mathrm{AB}}^{\mathrm{inel}}
\end{equation}
for the minimum bias collisions.
%%%%%%%%%%%%%%%%%%%%%%%%%%%%%%%%%%%%%%%%%%%%%%%%%%%%%%%%%%%%%%%%%%%%%
%%% Eq.~(\ref{BBC}) is exact at all orders of the multiply scattering
%%% theory.
%%% We cannot say something definite about the shape of
%%% $\xi_{\mathrm{AB}}$ for arbitrary $x$ but we may expect that it
%%% is a smooth function varying between the above extremes.
%%%%%%%%%%%%%%%%%%%%%%%%%%%%%%%%%%%%%%%%%%%%%%%%%%%%%%%%%%%%%%%%%%%%%

To use (\ref{AB->cX}) in a cascade calculation one should take into
account that nucleons and mesons are effectively produced in
nuclear collisions of different kind. Namely, the contribution from
central collisions is almost inessential for the nucleon component
of the cascade but quite important for light meson production.
Thus one can expect that effectively
$\xi_{\mathrm{AB}}^{\pi,K}>\xi_{\mathrm{AB}}^{p,n}$.
%%%%%%%%%%%%%%%%%%%%%%%%%%%%%%%%%%%%%%%%%%%%%%%%%%%%%%%%%%%%%%%%%%%%%
%%% In fact, experiments on light nuclei
%%% collisions \cite{IntNuclei} show that
%%% relation~(\ref{BBC}) is not valid for pion production.
%%%%%%%%%%%%%%%%%%%%%%%%%%%%%%%%%%%%%%%%%%%%%%%%%%%%%%%%%%%%%%%%%%%%%

From now on we use relation~(\ref{BBC}) for nucleon production by any
nucleus while for meson production we put
$\xi_{\mathrm{He-Air}}^{\pi,K}=\xi$ and vary the parameter $\xi$ from
0.517 (the BBC value) to 0.710 (an experimental upper limit derived
from the data on interactions of $\alpha$ particles with light nuclei
\cite{IntNuclei}). Similar (within $\pm15$\%) variations of the
parameters $\xi_{\mathrm{A-Air}}^{\pi,K}$ for other nuclear groups
lead to very small effects on the muon and neutrino fluxes; we do not
discuss these effects below.

It is evident that the SM corresponds to the formal limit
$\xi_{\mathrm{AB}}^c=1$ in eq.~(\ref{AB->cX}), if one simultaneously
puts $\sigma_{\mathrm{AB}}^{\mathrm{inel}}=
\sigma_{N\mathrm{B}}^{\mathrm{inel}}$ in the transport equations for
the nuclei and mesons. Below, we use this approximation making
efforts to represent the results of the Bartol code \cite{Bartol}
which is based on the TARGET model for meson production and the SM
for collisions of nuclei.

%%%%%%%%%%%%%%%%%%%%%%%%%%%%%%%%%%%%%%%%%%%%%%%%%%%%%%%%%%%%%%%%%%%%%
\protect\section{Numerical results: muon fluxes}
\label{sec:Muons}
%%%%%%%%%%%%%%%%%%%%%%%%%%%%%%%%%%%%%%%%%%%%%%%%%%%%%%%%%%%%%%%%%%%%%

In order to test the validity of our description of hadronic
interactions and shower development, we compare in this section some
numerical results for muon fluxes with data from several recent
balloon-borne experiments. Moreover, in order to evaluate
quantitatively theoretical uncertainties, we compare the calculations
obtained by using different codes and/or interaction models. Namely,
we consider numerical results obtained with:\\
{\em i)}   the CORT code, using our ``standard'' (KM+SS) interaction
           model; \\
{\em ii)}  the CORT code, using the TARGET model for $\pi/K$ meson
           production%
           \footnote{Due to technical reasons it is difficult to
                    adopt also the TARGET model for nucleon
                    production; we thus describe the nucleon
                    production by using the KM model everywhere.}
           (below, we refer to this calculation as ``CORT+TARGET'').
           \\
{\em iii)} the original Bartol Monte Carlo code \cite{Bartol} with
           the same (BESS+JACEE) fit for the primary spectra
           (``TARGET'').

%%%%%%%%%%%%%%%%%%%%%%%%%%%%%%%%%%%%%%%%%%%%%%%%%%%%%%%%%%%%%%%%%%%%%
%%% In this paper we shall limit ourselves by considering the muon
%%% data obtained during the periods close to solar minima and the
%%% data weakly affected by the solar activity (muon fluxes at large
%%% enough atmospheric depth and/or energies). The effects of solar
%%% modulation will be considered elsewhere.
%%%%%%%%%%%%%%%%%%%%%%%%%%%%%%%%%%%%%%%%%%%%%%%%%%%%%%%%%%%%%%%%%%%%%

\begin{figure}[htb]
\center\mbox{\epsfig{file=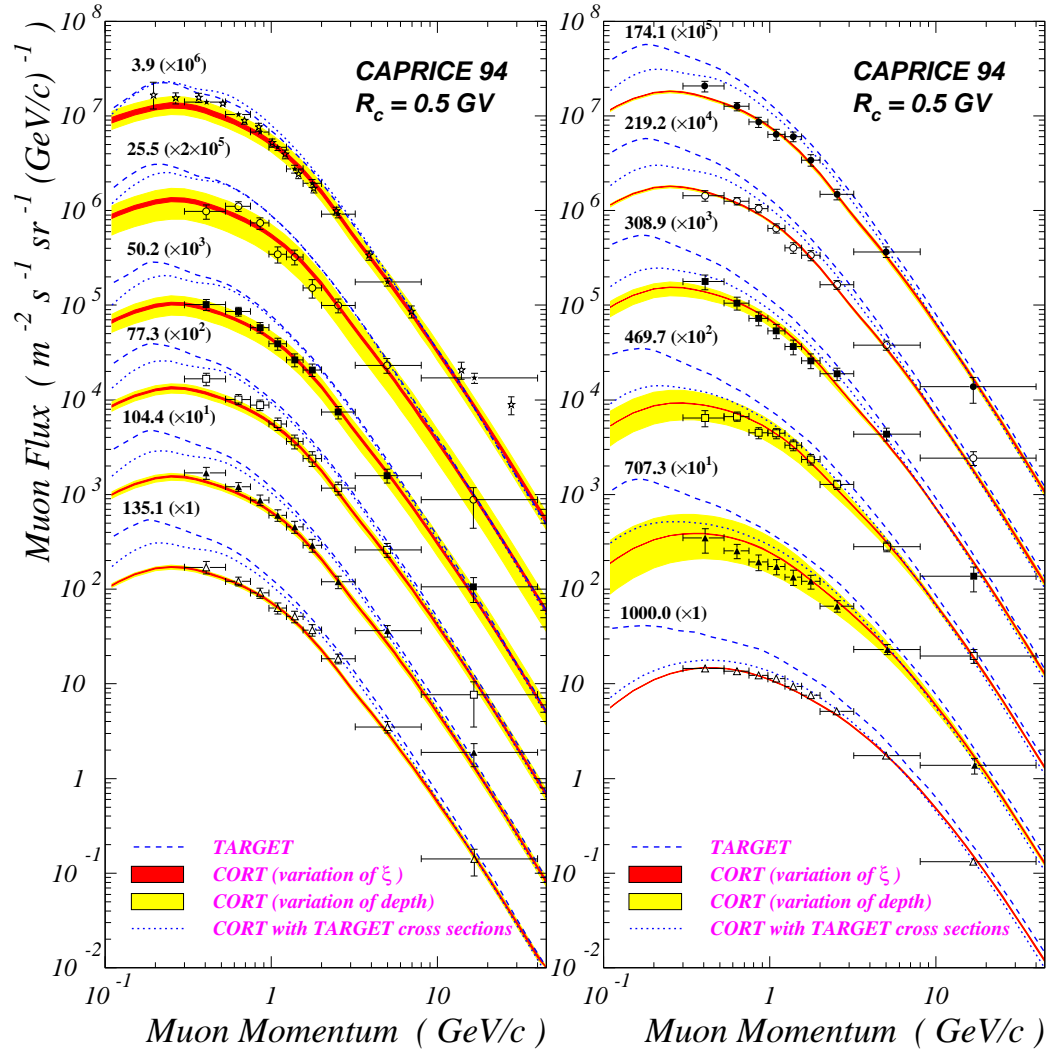,width=13.92cm}}
\protect\caption{Differential momentum spectra of negative muons
                 for different atmospheric depths. The data points
                 are from the balloon-borne experiment
                 CAPRICE\,94 \cite{Boezio99-00}. The curves and
                 filled areas are the results of calculations with
                 CORT and TARGET (see text). The numbers over the
                 curves indicate the flux-weighted average
                 atmospheric depths (in g/cm$^2$) and scale factors.
\label{f:CAPRICE94}}
\end{figure}

In fig.~\ref{f:CAPRICE94} we compare the $\mu^{-}$ fluxes calculated
with the three mentioned models for twelve atmospheric ranges with
the data of the balloon-borne experiment CAPRICE\,94
\cite{Boezio99-00}.
The measurements were performed during ground runs in Lynn Lake
($h=10^3$~g/cm$^2$) and during the continuous ascension of the
balloon to its float altitude ($h\approx4$~g/cm$^2$). The nominal
geomagnetic cutoff rigidity $R_c$ was about 0.5~GV and the detection
cone opening was of about $20^\circ$ around the vertical direction
with the typical zenith angle of $9^\circ$. The values of depths
indicated in fig.~\ref{f:CAPRICE94} are the Flux-weighted Average
Depths (FAD) \cite{Boezio99-00} for eleven atmospheric ranges
$\Delta h_i$ ($h_i<10^3$~g/cm$^2$).%
\footnote{There are also small (less than 3\%) spreads of the FADs
          within some of the ranges $\Delta h_i$, but these are
          completely negligible.}
%--------------------------------------------------------------------
%--------------------------------------------------------------------

Figure~\ref{f:CAPRICE94} contains many data. In order to make it
easily readable, let us explain the notation in detail.
The {\em solid curves} represent the results of our calculations for
each FAD.  They are obtained with the standard CORT and with $\xi$
varying between the limits mentioned in
section~\ref{sec:interactions}, that is within $\pm15$\%.
The thickness of the curves reflects the indetermination in the
parameter $\xi$.  For obvious reasons, the muon flux uncertainty due
to this indetermination is maximal at the top of the atmosphere (it
is about 15\% at $h=3.9$~g/cm$^2$ and $p=100$~MeV/c) and becomes
almost negligible for $h\gtrsim100$~g/cm$^2$.  The {\em wide filled
areas} display the variations of the muon fluxes inside the ranges
$\Delta h_i$.
%%%%%%%%%%%%%%%%%%%%%%%%%%%%%%%%%%%%%%%%%%%%%%%%%%%%%%%%%%%%%%%%%%%%%
%%% Namely, they are obtained considering the minimal and maximal
%%% muon fluxes within each ranges $\Delta h_i$.
%%%%%%%%%%%%%%%%%%%%%%%%%%%%%%%%%%%%%%%%%%%%%%%%%%%%%%%%%%%%%%%%%%%%%
It is important to note that the thickness of the bands is relatively
small just for the region of effective muon and neutrino production
that is in the neighborhood of the broad maxima of the muon flux
($100-300$~g/cm${}^2$). This means that, in this region, the
evaluation of the FAD cannot introduce any relevant uncertainty.
Outside the region of effective production of leptons, the amplitude
of the muon flux variations increases with decreasing muon momenta on
account for the strong dependence of the meson production rate upon
the depth at $h\lesssim100$~g/cm${}^2$ and the growing role of the
muon energy loss and decay at $h\gtrsim300$~g/cm${}^2$.  The {\em
dashed curves} show the results obtained using the original Bartol
code and the {\em dotted curves} are the results of the CORT+TARGET
model (we shall comment them later).

%%%%%%%%%%%%%%%%%%%%%%%%%%%%%%%%%%%%%%%%%%%%%%%%%%%%%%%%%%%%%%%%%%%%%
%%% As was shown in ref. \cite{Naumov93-00}, the KM model
%%% satisfactory describe available data on secondary proton and
%%% neutron fluxes. However, these data are too scanty for reliable
%%% conclusions.
%%%%%%%%%%%%%%%%%%%%%%%%%%%%%%%%%%%%%%%%%%%%%%%%%%%%%%%%%%%%%%%%%%%%%

\begin{figure}[t]
\center\mbox{\epsfig{file=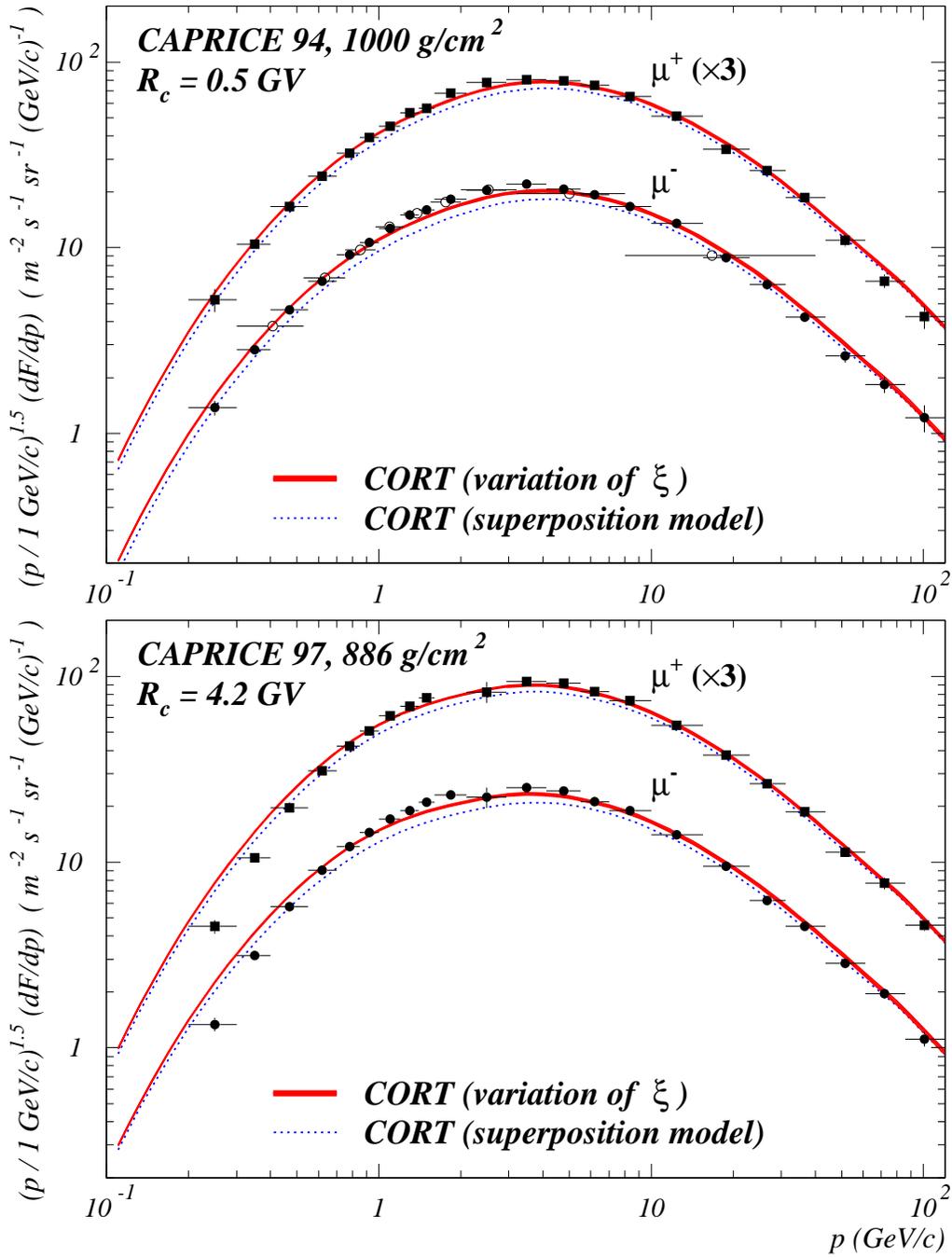,width=13.62cm}}
\protect\caption{Differential momentum spectra of positive and
                 negative muons at the ground level. The data points
                 are from the experiments
                 CAPRICE\,94 \cite{Boezio99-00,Kremer99} and
                 CAPRICE\,97 \cite{Kremer99}.
                 The curves are calculated with CORT.
\label{f:CAPRICE_gl}}
\end{figure}

Figure~\ref{f:CAPRICE_gl} shows a comparison of the calculated
differential momentum spectra of $\mu^+$ and $\mu^-$ with the ground
level data obtained in the experiments CAPRICE\,94
\cite{Boezio99-00,Kremer99} ($h=1000$~g/cm$^2$, $R_c=0.5$~GV) and
CAPRICE\,97 \cite{Kremer99} ($h=886$~g/cm$^2$, $R_c=4.2$~GV).  The
detection cone in both experiments was the same as described above.
The calculation are done using the KM+SS interaction model. Variation
of $\xi$ leads to a small ($\lesssim3$\%) effect. One sees quite a
good agreement between the theory and data.

For comparison, the result obtained using the SM is also shown. The
data below $\sim10$~GeV/c are precise enough in order to conclude
that they disfavor the superposition model. Let us note that the
calculated sea-level ($h=10^3$~g/cm$^2$) $\mu^++\mu^-$ flux is
systematically lower than that obtained using the old version of CORT
(see ref. \cite{Bugaev98}): the difference is about 18\% for
$p=10$~GeV/c and vanishes only at $p\gtrsim80$~GeV/c. The main reason
of this difference is that the BESS+JACEE primary spectrum is
essentially lower than the spectrum used in ref. \cite{Bugaev98} and
in the earlier calculations \cite{BN1,BN2} for energies below
$\sim200$~GeV/nucleon. Another reasons pertain to the changes in the
inclusive cross sections and improved description of muon
propagation.

\begin{figure}[htb]
\centering\mbox{\epsfig{file=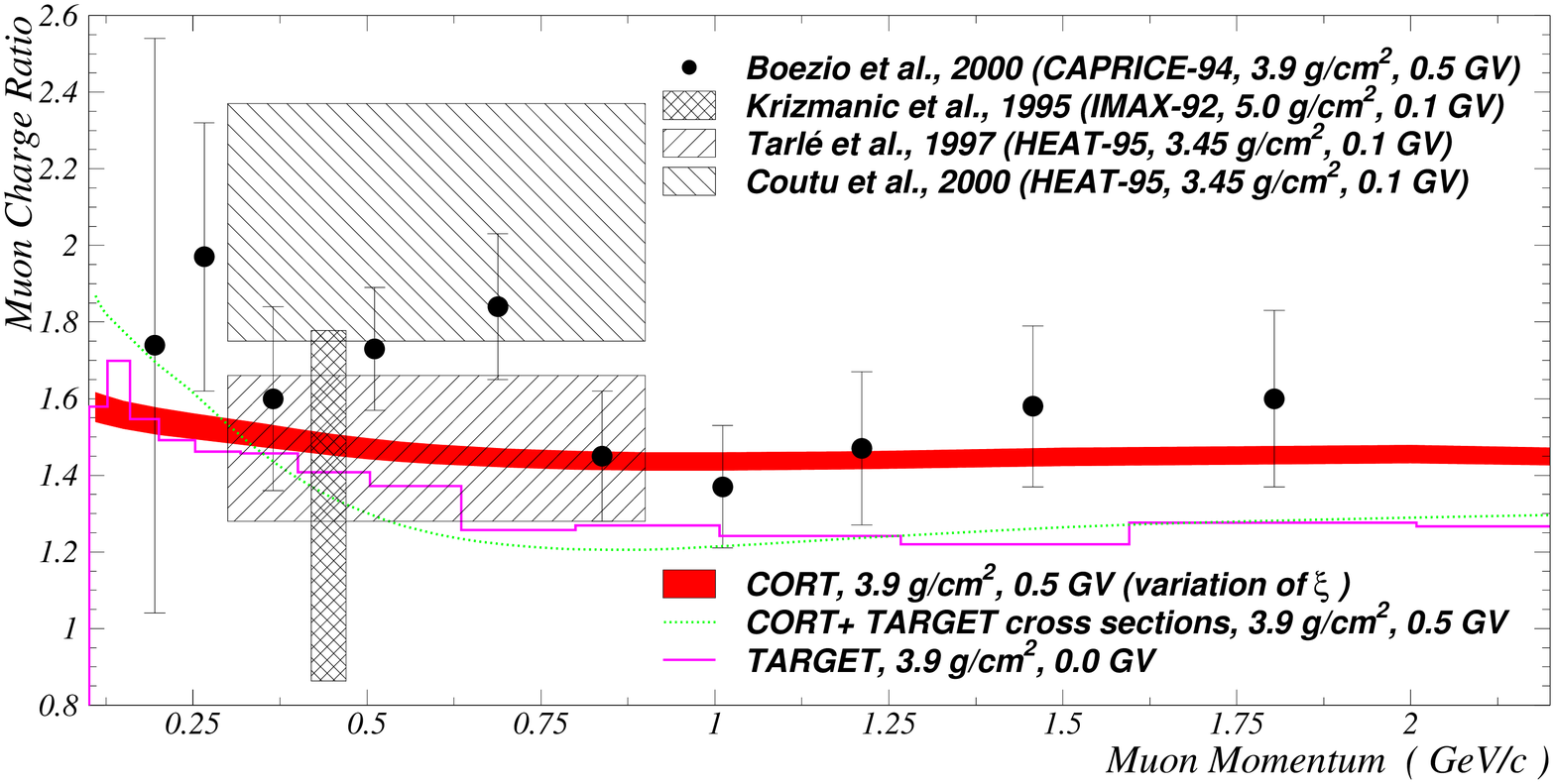,width=13.35cm}}
\centering\mbox{\epsfig{file=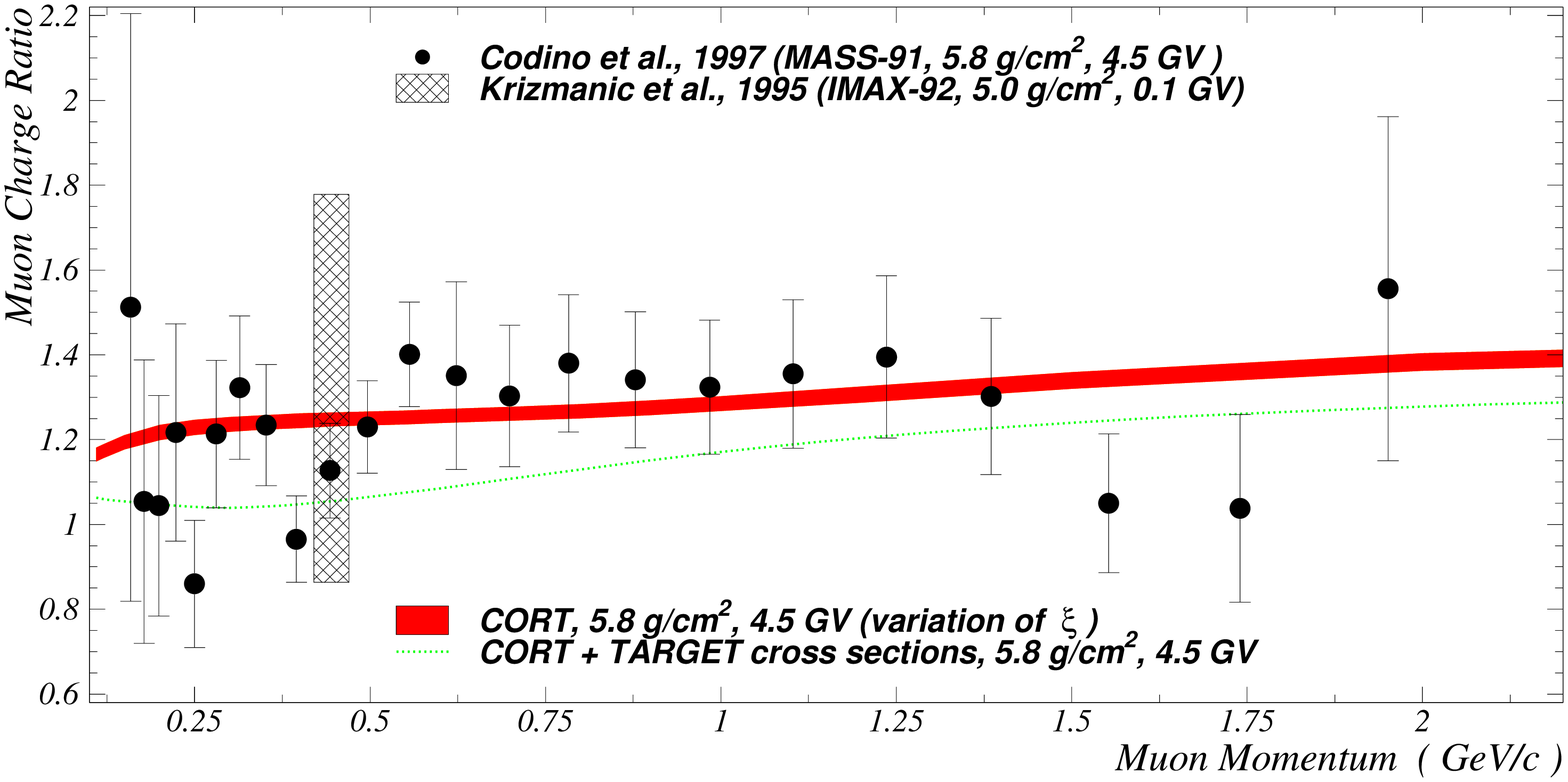,width=13.35cm}}
\protect\caption{Muon charge ratio near the top of the atmosphere.
                 The points and shaded rectangles represent
                 the data from the experiments
                 MASS\,91 \cite{Codino97},
                 IMAX\,92 \cite{Krizmanic95},
                 HEAT\,95 \cite{Tarle97,Coutu00} and
                 CAPRICE\,94 \cite{Boezio99-00}.
                 The curves and filled areas are the results of
                 calculations with CORT and TARGET.
\label{f:CRtop}}
\end{figure}

\begin{figure}[htb]
\centering\mbox{\epsfig{file=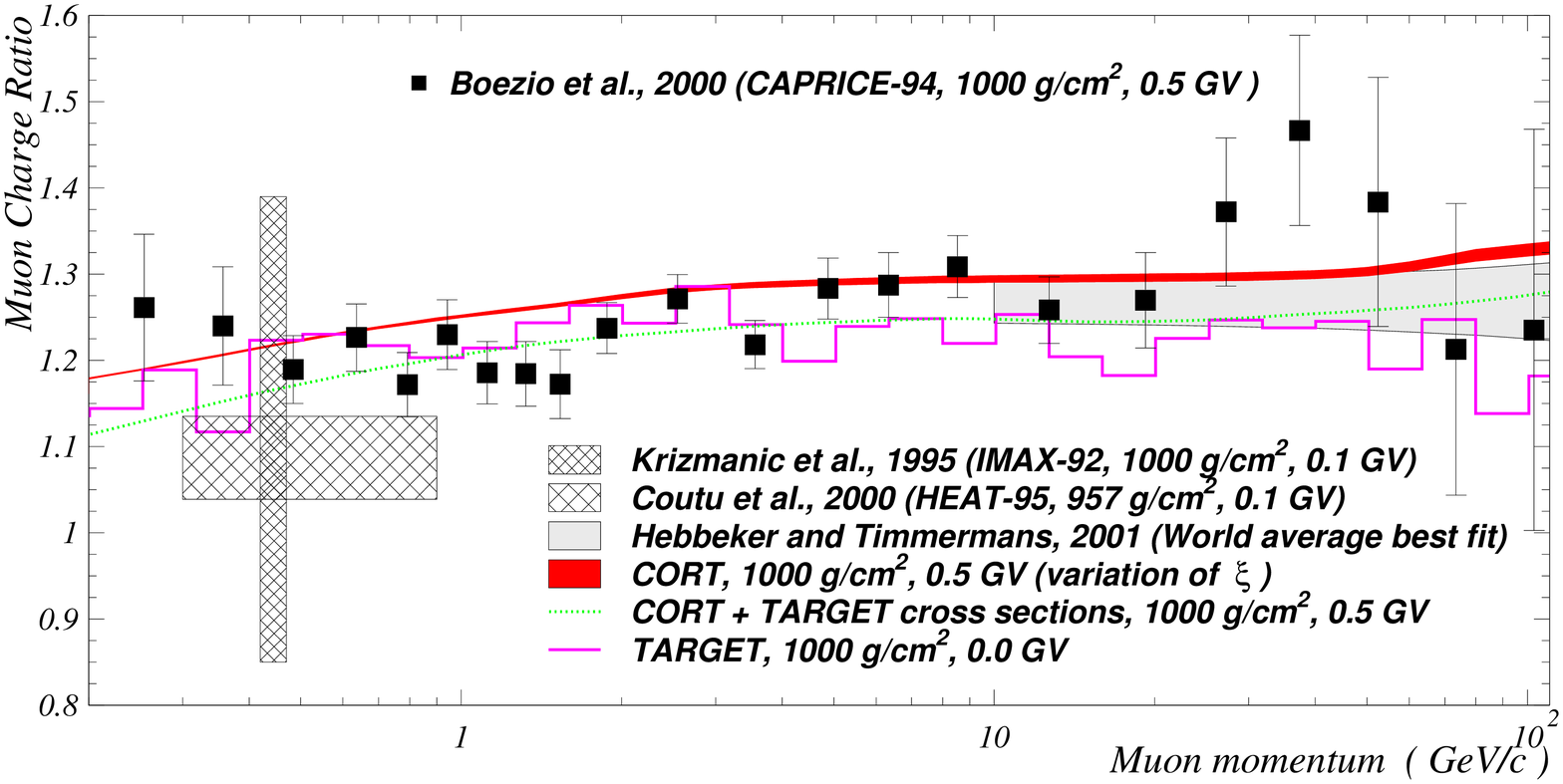,width=13.35cm}}
\centering\mbox{\epsfig{file=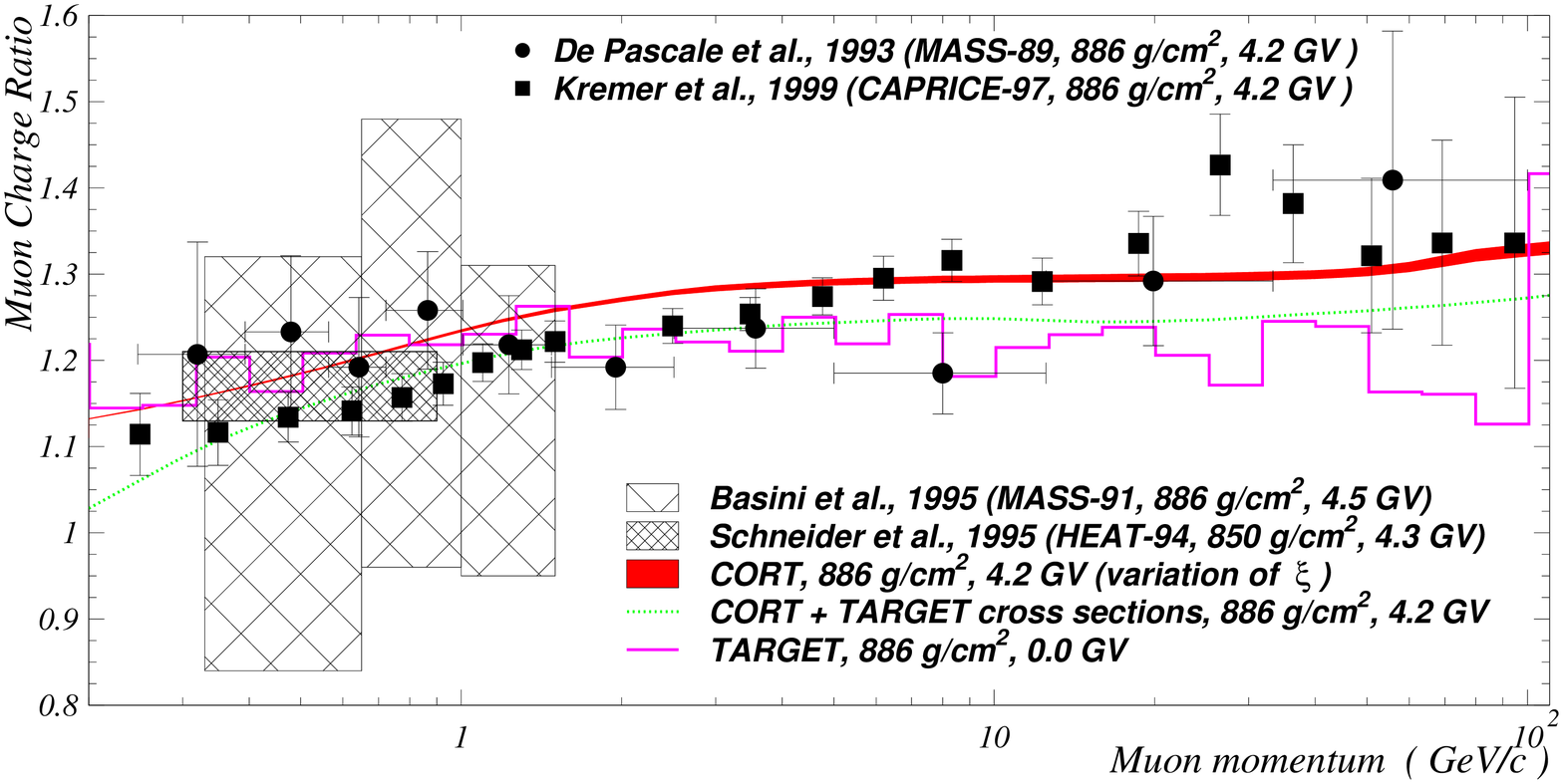,width=13.35cm}}
\protect\caption{Muon charge ratio at the ground level. The points
                 and shaded rectangles represent the data from
                 the experiments
                 MASS\,89 \cite{DePascale93},
                 MASS\,91 \cite{Basini95},
                 IMAX\,92 \cite{Krizmanic95},
                 HEAT\,94 \cite{Schneider95},
                 HEAT\,95 \cite{Coutu00},
                 CAPRICE\,94 \cite{Boezio99-00} and
                 CAPRICE\,97 \cite{Kremer99}.
                 The curves and filled areas represent the results of
                 calculations with CORT and TARGET and the
                 best fit obtained in ref. \cite{Hebbeker01}
                 from many experiments at $p\geq10$~GeV/c.
\label{f:CRgl}}
\end{figure}

In figs.~\ref{f:CRtop} and~\ref{f:CRgl} we present the muon charge
ratio $\mu^+/\mu^-$ as a function of muon momentum for different
atmospheric depths and geomagnetic cutoffs. The data are from several
balloon experiments:
MASS\,89 \cite{DePascale93},
MASS\,91 \cite{Basini95,Codino97},
IMAX\,92 \cite{Krizmanic95},
HEAT\,94 \cite{Schneider95},
HEAT\,95 (two data handlings of the same data sample given
in \cite{Tarle97} and \cite{Coutu00}),
CAPRICE\,94 \cite{Boezio99-00} and
CAPRICE\,97 \cite{Kremer99}.
The world average best fit, recently obtained in ref.
\cite{Hebbeker01} by an analysis of the data from many ground level
experiments at $p\geq10$~GeV/c, is also shown.  The muon charge
ratio, being sensitive to the primary chemical composition and to the
$\pi^+/\pi^-$ ratio at production, is a useful tool for testing the
predicted $\overline{\nu}/\nu$ ratio.

\clearpage
%--------------------------------------------------------------------
%--------------------------------------------------------------------
\clearpage
%--------------------------------------------------------------------
%--------------------------------------------------------------------
%--------------------------------------------------------------------
%--------------------------------------------------------------------

Several important conclusions can be deduced by comparing the various
data presented in figs. \ref{f:CAPRICE94}, \ref{f:CAPRICE_gl},
\ref{f:CRtop} and \ref{f:CRgl}.

\begin{itemize}
\item There is a substantial agreement between the ``standard'' CORT
      predictions and the current muon data within wide ranges of
      muon momenta and atmospheric depths. In particular, the
      agreement is good for the region of effective production
      of leptons, in which the spread of the data (partially related
      to the procedure of determination of the FAD and mean muon
      momenta) is minimal. This provides an evidence for the validity
      of our model.
%%%   description of hadronic interactions and shower development.
\item The comparison between CORT and CORT+TARGET allows to elucidate
      the ``hadronic'' uncertainties in muon fluxes.  The only
      difference between the two calculations is indeed the assumed
      meson production model and the model for nucleus-nucleus
      interactions. The $\mu^-$ fluxes obtained using the TARGET
      model systematically exceed the results obtained with our
      standard model. In order to have a quantitative estimate of the
      hadronic uncertainties, we note that the muon flux at
      $p=1$~GeV/c and $h=(100-300)$~g/cm${}^2$ (i.e. the range where
      muons and neutrinos are effectively produced) increases by
      about 50\% when the TARGET model is applied. For comparison, we
      remind that the accuracy of the CAPRICE\,94 experiment in this
      region is about 15\%.
\item We can see that, except for very small depths, there is a
      serious disagreement between the CORT+TARGET and the original
      Bartol results. The disagreement grows fast with decreasing
      muon momentum and with increasing depth, and it is comparable
      (or even larger) to the hadronic uncertainties.  Since the two
      calculations are obtained by using the same primary spectrum
      and the same meson production model, the observed difference
      can only be due to different features in the nuclear cascade
      development and muon propagation.  On the other hand, the
      observed agreement at floating altitude, where such features
      are not important, assures the correctness of our use of the
      TARGET meson production model. Possible sources for the
      observed difference could be nucleon elasticity distributions,
      meson regeneration processes, muon energy losses and some more
      tiny features of the cascade models under consideration.
\end{itemize}

%%%%%%%%%%%%%%%%%%%%%%%%%%%%%%%%%%%%%%%%%%%%%%%%%%%%%%%%%%%%%%%%%%%%%
\protect\section{Numerical results: neutrino fluxes}
\label{sec:Results}
%%%%%%%%%%%%%%%%%%%%%%%%%%%%%%%%%%%%%%%%%%%%%%%%%%%%%%%%%%%%%%%%%%%%%

In this section we present our numerical results for atmospheric
neutrino (AN) flux.
%%%%%%%%%%%%%%%%%%%%%%%%%%%%%%%%%%%%%%%%%%%%%%%%%%%%%%%%%%%%%%%%%%%%%
%% (see figures \ref{f:AN_K_A}, \ref{f:AN_K_ZD} and
%% tabels \ref{t:FluxRatios}, \ref{t:FlavorRatios}).
%% For simplicity, we consider only the case of (Super)Kamiokande.
%%%%%%%%%%%%%%%%%%%%%%%%%%%%%%%%%%%%%%%%%%%%%%%%%%%%%%%%%%%%%%%%%%%%%
Due to geomagnetic effects, the AN spectra and angular distributions
are quite different for different underground neutrino experiments.
However, for the aims of present study it is enough to consider only
one representative case. Let us confine ourselves to
results pertinent to the Kamioka site.

\begin{figure}[htb]
\centering\mbox{\epsfig{file=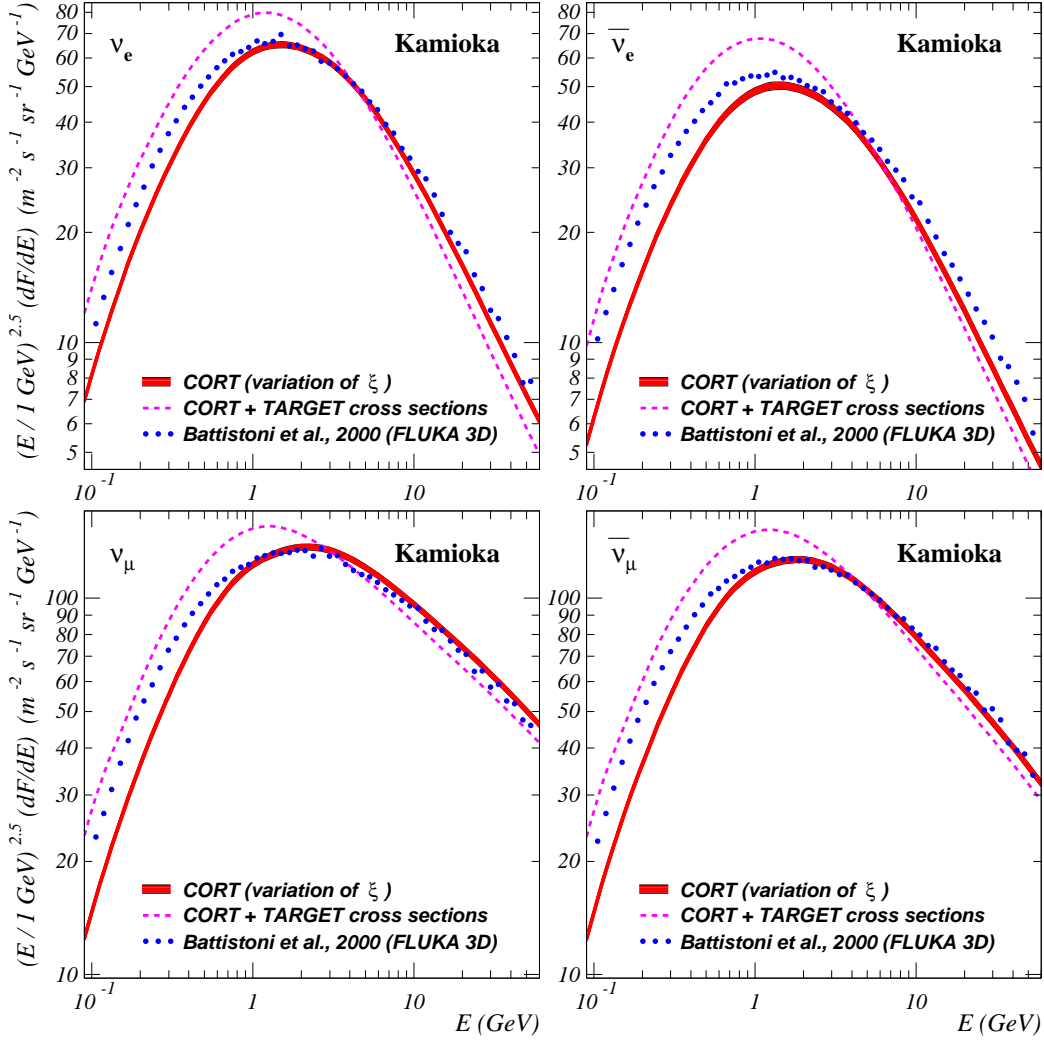,width=13.92cm}}
\protect\caption{Scaled $4\pi$ averaged fluxes of $\nu_e$,
                 $\overline{\nu}_e$, $\nu_\mu$ and
                 $\overline{\nu}_\mu$ for Kamioka site.
                 Shaded areas represent the result of CORT obtained
                 with $\xi$ varying between 0.517 and 0.710.
                 The dashed curves are the result of CORT obtained
                 with the TARGET model for meson production and
                 superposition model for collisions of nuclei.
                 The circles are for the result of a 3D calculation
                 based on the FLUKA code \cite{FLUKA}.
\label{f:AN_K_A}}
\end{figure}

\begin{figure}[htb]
\centering\mbox{\epsfig{file=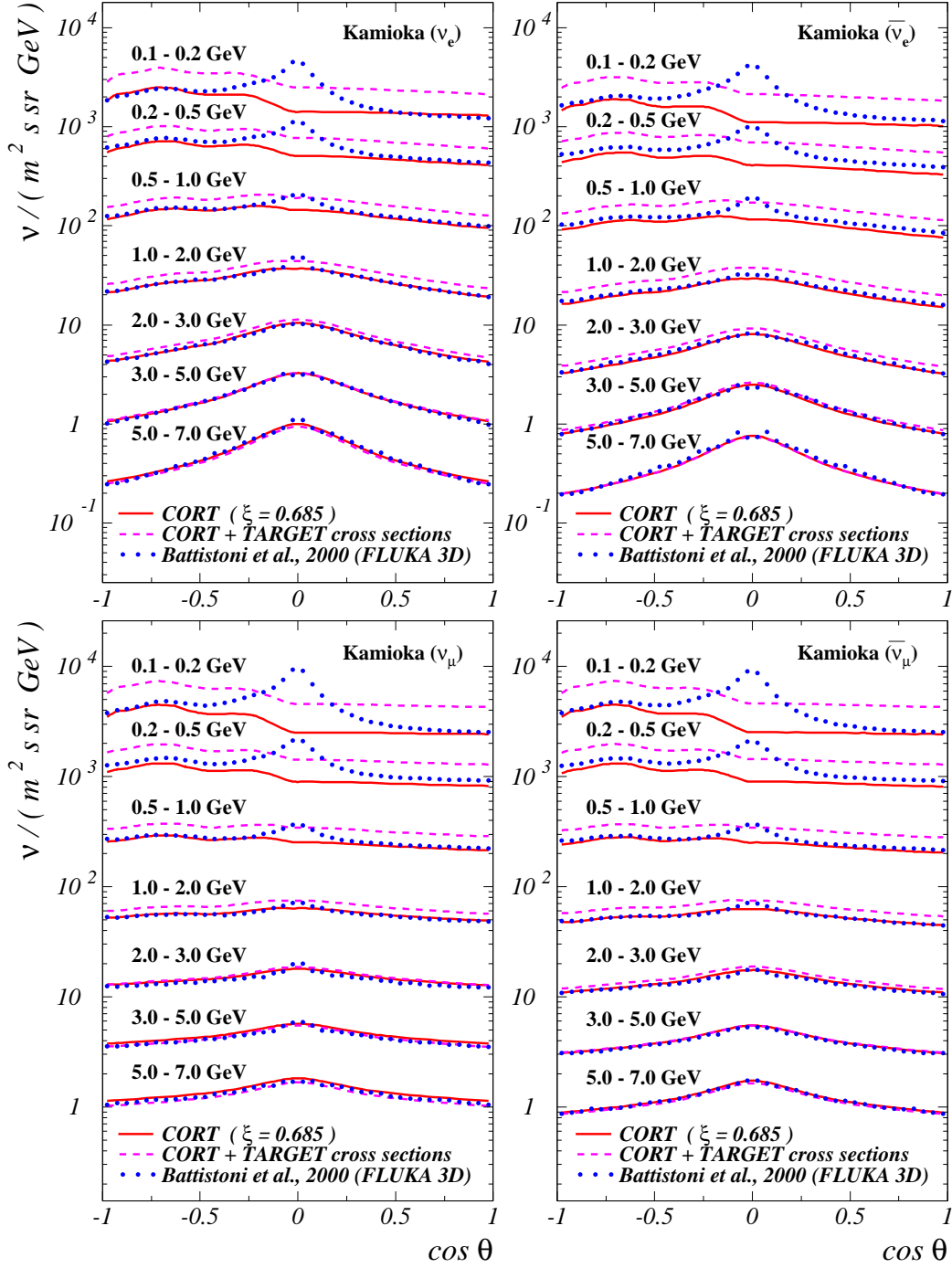,width=13.92cm}}
\protect\caption{Zenith angle distributions of $\nu_e$,
                 $\overline{\nu}_e$, $\nu_\mu$ and
                 $\overline{\nu}_\mu$ for several energy ranges.
                 Solid curves represent the result of CORT obtained
                 with $\xi=0.685$ and the rest notation is the same as
                 in fig.~\protect\ref{f:AN_K_A}.
\label{f:AN_K_ZD}}
\end{figure}

Figure~\ref{f:AN_K_A} shows the $\nu_e$, $\overline{\nu}_e$,
$\nu_\mu$ and $\overline{\nu}_\mu$ energy spectra averaged over all
zenith and azimuth angles.  The shaded areas are the results obtained
with CORT using our standard interaction model. The widths of the
areas indicate the uncertainty due to variations of the $\xi$
parameter. One sees that this uncertainty is at most 6\% and thus it
is negligible. The dashed curves correspond to the CORT+TARGET model,
while the circles show the results by a new (preliminary as yet) 3D
calculation based on the FLUKA Monte Carlo simulation package
\cite{FLUKA} (see also \cite{Battistoni01}).  Note that the primary
spectrum used in the latter calculation is also a parametrization of
the BESS\,98 data, which is however not identical to our BESS+JACEE
fit.
In fig.~\ref{f:AN_K_ZD} we compare the zenith-angle distributions of
$\nu_e$, $\overline{\nu}_e$, $\nu_\mu$ and $\overline{\nu}_\mu$,
calculated with CORT (adopting $\xi=0.685$), CORT+TARGET and FLUKA;
the distributions are averaged over azimuth angles and over seven
energy ranges. The comparison allows to ``highlight'' the 3D effects
which are highly dependent on neutrino energy and direction of
arrival.
%--------------------------------------------------------------------
%--------------------------------------------------------------------

%--------------------------------------------------------------------
%--------------------------------------------------------------------

As a first point, we note that our standard model leads to the lowest
AN flux at low energies. Our flux is also essentially lower than that
predicted by Bartol group \cite{Bartol} and by Honda et al.
\cite{Honda}%
\footnote{Let us remind that the results of refs. \cite{Bartol,Honda}
          were obtained with different (higher) fluxes of the primary
          cosmic rays.}
and that are used in many analyses of the data from IMB, Fr\'ejus,
NUSEX, Kamiokande, Super-Kamiokande, SOUDAN\,2 and MACRO.

We do not enter here in the complex problem of the interpretation of
the AN anomaly in the light of our result. We want to remark,
however, that, in the context of our model, it is difficult to
increase the AN flux without spoiling the agreement with the current
data on hadronic interactions, primary spectrum and muon fluxes.
Let us emphasize that our low AN flux, when it is applied to the
analysis of the underground neutrino experiments, results in some
electron excess together with (or rather than) the muon deficit in
the neutrino induced events.

%--------------------------------------------------------------------
\begin{table}[t]
\centering
\protect\caption{$4\pi$ averaged AN fluxes for Kamioka site
                 calculated with CORT+TARGET and FLUKA and
                 normalized to CORT.}
\label{t:FluxRatios}
\vspace{3mm}
\begin{tabular}{c|cccc|cccc}
\hline
$\Delta E_\nu$ & \multicolumn{4}{c|}{CORT+TARGET}
               & \multicolumn{4}{c }{FLUKA}                       \\
  (GeV)        &
 $\nu_e$ & $\overline{\nu}_e$ & $\nu_\mu$ & $\overline{\nu}_\mu$ &
 $\nu_e$ & $\overline{\nu}_e$ & $\nu_\mu$ & $\overline{\nu}_\mu$  \\
\hline
 0.1--0.2 & 1.64 & 1.78 & 1.72 & 1.73 & 1.25 & 1.45 & 1.41 & 1.40 \\
 0.2--0.3 & 1.51 & 1.68 & 1.60 & 1.61 & 1.23 & 1.38 & 1.34 & 1.33 \\
 0.3--0.5 & 1.42 & 1.61 & 1.48 & 1.50 & 1.17 & 1.28 & 1.22 & 1.22 \\
 0.5--0.7 & 1.34 & 1.50 & 1.36 & 1.38 & 1.10 & 1.19 & 1.11 & 1.12 \\
 0.7--1.0 & 1.28 & 1.41 & 1.27 & 1.30 & 1.05 & 1.12 & 1.04 & 1.05 \\
 1.0--2.0 & 1.20 & 1.30 & 1.17 & 1.20 & 1.02 & 1.05 & 0.99 & 1.01 \\
 2.0--3.0 & 1.10 & 1.17 & 1.03 & 1.08 & 0.98 & 1.03 & 0.96 & 0.97 \\
 3.0--5.0 & 1.02 & 1.07 & 0.95 & 1.01 & 0.99 & 1.02 & 0.95 & 0.97 \\
 5.0--7.0 & 0.95 & 0.99 & 0.91 & 0.96 & 1.01 & 1.05 & 0.94 & 0.99 \\
 7.0--10. & 0.91 & 0.95 & 0.89 & 0.93 & 1.03 & 1.09 & 0.95 & 1.04 \\
 10.--20. & 0.87 & 0.91 & 0.88 & 0.91 & 1.07 & 1.12 & 0.96 & 1.02 \\
%20.--30. & 0.83 & 0.87 & 0.87 & 0.89 & 1.10 & 1.15 & 0.94 & 1.02 \\
\hline
\end{tabular}
\vspace{5mm}
\end{table}
%--------------------------------------------------------------------

Let us now discuss in detail the differences between our results
and those obtained within CORT+TARGET and FLUKA. The quantitative
comparison is given in tables~\ref{t:FluxRatios} and
\ref{t:FlavorRatios}. Table~\ref{t:FluxRatios} shows the ratios of
$4\pi$ averaged AN fluxes, obtained with CORT+TARGET and FLUKA to
those with CORT for several energy ranges.
In table~\ref{t:FlavorRatios} we tabulate the ``flavor ratio'',
defined by
\[
R_\nu=\left(\nu_e  +\frac{1}{3}\overline{\nu}_e  \right)\big/
      \left(\nu_\mu+\frac{1}{3}\overline{\nu}_\mu\right),
\]
where $\nu_e$, $\overline{\nu}_e$, etc. stand for the $4\pi$ averaged
fluxes. This quantity is representative for the ratio of $e$ like to
$\mu$ like single-ring contained events measured in water Cherenkov
detectors and to ``showers-to-tracks'' ratio measured in iron
detectors.
%--------------------------------------------------------------------
\begin{table}[htb]
\centering
\protect\caption{Neutrino flavor ratios $R_\nu$ calculated with CORT,
                 CORT+TARGET and FLUKA for Kamioka site.}
\label{t:FlavorRatios}
\vspace{3mm}
\begin{tabular}{c|ccc}
\hline
$\Delta E_\nu$ (GeV) & CORT & CORT+TARGET & FLUKA \\ \hline
 0.1--0.2            & 0.52 &    0.51     & 0.48  \\
 0.2--0.3            & 0.52 &    0.50     & 0.49  \\
 0.3--0.5            & 0.51 &    0.50     & 0.50  \\
 0.5--0.7            & 0.50 &    0.50     & 0.50  \\
 0.7--1.0            & 0.49 &    0.50     & 0.50  \\
 1.0--2.0            & 0.47 &    0.49     & 0.49  \\
 2.0--3.0            & 0.44 &    0.47     & 0.45  \\
 3.0--5.0            & 0.40 &    0.43     & 0.42  \\
 5.0--7.0            & 0.36 &    0.37     & 0.38  \\
 7.0--10.            & 0.32 &    0.32     & 0.34  \\
 10.--20.            & 0.26 &    0.26     & 0.29  \\
%20.--30.            & 0.20 &    0.19     & 0.23  \\
\hline
\end{tabular}
\end{table}
%--------------------------------------------------------------------

The comparison between CORT and CORT+TARGET predictions provides a
direct determination of the ``hadronic'' uncertainties in the AN
fluxes. We remind indeed that these two results are obtained by using
the same primary fluxes and the same code; thus the observed
differences is only due to the assumed meson production model.  At
low energies, the CORT+TARGET predictions drastically exceed those
from CORT and even at energies about (2-2.5)~GeV one can observe a
sizeable difference between the two results.  At $E_\nu\simeq0.5$~GeV
(tab.~\ref{t:FluxRatios}), the discrepancy is of the order of
40-50\%.  This shows that the hadronic uncertainties are quite
relevant in the energy range which is responsible for the sub-GeV
event rate in the Super-Kamiokande detector.

Figs.~\ref{f:AN_K_A} and~\ref{f:AN_K_ZD} show that CORT and FLUKA
neutrino fluxes are in good agreement for any zenith angle at
energies above $\sim1$~GeV. On the other hand, CORT is systematically
lower than FLUKA for $E_\nu\lesssim1$~GeV. At the energy
$E_\nu=0.5$~GeV the FLUKA $4\pi$ averaged $\nu_e$,
$\overline{\nu}_e$, $\nu_\mu$ and $\overline{\nu}_\mu$ fluxes exceed
the corresponding CORT results by about 15\%, 27\%, 18\% and 18\%,
respectively. This discrepancy is only partially due to 3D effects
(fig.\ref{f:AN_K_ZD}), which can account for an increase
$\lesssim10\%$, and it is most probably related to differences between
the FLUKA and CORT hadronic interaction models.

Finally, as it is seen from table~\ref{t:FlavorRatios}, the flavor
ratio is essentially the same in all three models, except for very
low energies. As one can conclude from fig.~\ref{f:AN_K_ZD}, the same
is also true for the ``up-to-down'' asymmetry, that is for the ratio
of upward-going ($\cos\theta<0$) and downward-going ($\cos\theta>0$)
neutrino fluxes.

%%%%%%%%%%%%%%%%%%%%%%%%%%%%%%%%%%%%%%%%%%%%%%%%%%%%%%%%%%%%%%%%%%%%%
\protect\section{Conclusions}
\label{sec:Conclusions}
%%%%%%%%%%%%%%%%%%%%%%%%%%%%%%%%%%%%%%%%%%%%%%%%%%%%%%%%%%%%%%%%%%%%%

Let us summarize here the main points of our discussion.

\begin{itemize}
\item
Calculations with a new version of the 1D code CORT, using the
updated KM+SS interaction model and the BESS+JACEE primary spectrum
are in agreement with the recent data on muon momentum spectra and
charge ratios measured at different atmospheric depths and
geomagnetic locations.
\item
The uncertainty due to indetermination of the
parameters of our model for nucleus-nucleus interactions is
comparatively small everywhere, except for high altitudes.
Some experimental uncertainties (like those due to determination of
the flux-weighted average depths) are negligible for the regions of
effective generation of leptons and at ground level. Therefore the
agreement between the predictions and the muon data obtained at these
depths provides a conclusive evidence for validity of our approach.
\item
Our atmospheric neutrino flux is systematically lower than those used
in the current analyses of the data from underground experiments,
while the neutrino flavor ratio and ``up-to-down'' asymmetry are
essentially the same as in all recent calculations.
%%% except those from ref. \cite{Plyaskin01}.
We stress that, in the context of our model, it is difficult to
increase the AN flux without spoiling the agreement with the current
data on hadronic interactions, primary spectrum and muon fluxes.

\end{itemize}

In order to estimate theoretical uncertainties in the muon and
neutrino fluxes we compared our results with those obtained by
using different codes and/or interaction models, with the following
conclusions.

\begin{itemize}
\item By implementing {\em two different interaction models within
      the same code}, we have shown that the ``hadronic''
      uncertainties are quite large at low energies for both muon and
      neutrino fluxes.  Specifically, by using the TARGET meson
      production model instead of the KM+SS, we obtained a sizeable
      increase of neutrino fluxes at energies up to
      $E_\nu\simeq3$~GeV.  At the energy $E_\nu\simeq0.5$~GeV
      representative for the sub-GeV events in the Super-Kamiokande
      detector, one observes a 40-50\% increase of the AN fluxes.
\item By comparing {\em two different codes implementing the same
      interaction model} (TARGET), we have shown that other sources
      of theoretical uncertainties are also relevant. Namely, a
      sizeable difference exists between the muon fluxes predicted by
      the original Bartol code and those obtained by implementing the
      TARGET interaction model within the CORT code. We believe that
      the observed disagreement is due to the differences in the
      calculation of nuclear cascade, muon energy loss and decay.
\item
A comparison of the $4\pi$ averaged AN fluxes calculated with CORT
and FLUKA shows rather good agreement at $E_\nu>0.7-0.8$~GeV
(the difference is typically less than 15\%) and a sizable
disagreement at lower energies. A comparison of neutrino zenith-angle
distributions, predicted by CORT and FLUKA, suggests that the main
sources of the disagreement at low energies are the differences
in the adopted interaction models and 3D effects.
\end{itemize}

Finally we conclude that the AN spectra and angular distributions
predicted with CORT using the BESS+JACEE primary spectrum and KM-SS
interaction model are quite safe at energies above $\sim1$~GeV while
the low-energy range (which is of special importance, as a source of
background to the proton decay experiments) is still uncertain.

\ack %\acknowledgments

We wish to acknowledge helpful discussions with G.~Battistoni,
M.~Boezio, M.~Circella and S.~Coutu. We also thank them for providing
us with their unpublished results. We are grateful to T.~Gaisser and
T.~Stanev for making available to us their TARGET Monte Carlo
package.
%%%%%%%%%%%%%%%%%%%%%%%%%%%%%%%%%%%%%%%%%%%%%%%%%%%%%%%%%%%%%%%%%%%%%
%%% The work of V.\,N. was supported in part by the Ministry of
%%% Education of Russian Federation, within the framework of the
%%% program ``Universities of Russia -- Basic Researches''.
%%%%%%%%%%%%%%%%%%%%%%%%%%%%%%%%%%%%%%%%%%%%%%%%%%%%%%%%%%%%%%%%%%%%%

%\input{Adds/CORT_A.TeX}

\end{document}